\documentclass[smallextended]{svjour3}       
\smartqed  
\usepackage{graphicx}

\usepackage{epsf,epsfig,graphicx}
\usepackage[nottoc, notlof, notlot]{tocbibind}

\usepackage{array}
\usepackage{units}
\usepackage{textcomp}
\usepackage{amsmath}
\usepackage{amssymb}
\usepackage{esint}
\usepackage{amsfonts}
\usepackage{multirow}
\usepackage{rotating}
\usepackage{paralist}

\usepackage{multicol}
\usepackage{hyperref}

\usepackage{subcaption}

\usepackage{wrapfig}
\usepackage{sidecap}
\usepackage{color}
\usepackage{caption}
\usepackage{paralist}
\usepackage{lscape}
\newcommand{\be}{\begin{equation}}
\newcommand{\ee}{\end{equation}}

\def \eam{ASTROMEV }
\def \eamp{ASTROMEV}

\def \mt{ }

\usepackage{aas_macros}

\begin{document}

\title{Gamma-ray Astrophysics in the MeV Range
}
\subtitle{The ASTROGAM Concept and Beyond}


\author{Alessandro De Angelis         \and
        Vincent Tatischeff  \and
               Andrea Argan \and
                        S{\o}ren Brandt \and
                                Andrea Bulgarelli \and
                                        Andrei Bykov \and
                                                Elisa Costantini  \and
                                                        Rui Curado da Silva \and
                                                                Isabelle A. Grenier \and
                                                                        Lorraine Hanlon \and
                                                                                Dieter Hartmann \and
                                                                                        Margarida Hernanz   \and
               Gottfried Kanbach  \and
                        Irfan Kuvvetli \and
                                Philippe Laurent  \and
                                        Mario N. Mazziotta \and
                                                Julie McEnery \and
                                                       Aldo Morselli   \and
                                                                Kazuhiro Nakazawa \and  Uwe Oberlack  \and
                                                                        Mark Pearce  \and
                                                                                Javier Rico \and
                        Marco Tavani \and
                                Peter von Ballmoos \and
                                       Roland Walter   \and
                                                Xin Wu  \and
                                                       Silvia Zane  \and
                                                                Andrzej  Zdziarski  \and
                                                                        Andreas Zoglauer
}


\institute{A. De Angelis \at
             Dipartimento di Fisica e Astronomia ``Galileo Galilei''\\
\noindent via Marzolo 8, Padova I-35131, Italy\\
              Tel.: +39-049-827-5942\\
              \email{alessandro.deangelis@unipd.it}           
           \and
           V. Tatischeff \at
Universit\'e Paris-Saclay, CNRS/IN2P3, IJCLab,\\
 F-91405, Orsay, France
}

\date{Received: date / Accepted: date}

\maketitle

\begin{abstract}
The energy range between about 100 keV and 1 GeV is of interest for a vast class of astrophysical topics. In particular, (1) it is the missing ingredient for understanding extreme processes in the multi-messenger era; (2) it allows localizing cosmic-ray interactions with background material and radiation in the Universe, and spotting the reprocessing of these particles; (3) last but not least, gamma-ray emission lines trace the formation of elements in the Galaxy and beyond. In addition, studying the still largely unexplored MeV domain of astronomy would provide for a rich observatory science, including the study of compact objects, solar- and Earth-science, as well as fundamental physics. The technological development of silicon microstrip detectors makes it possible now to detect MeV photons in space with high efficiency and low background. During the last decade, a concept of detector (``ASTROGAM") has been proposed to fulfil these goals, based on a silicon hodoscope, a 3D position-sensitive calorimeter, and an anticoincidence detector. In this paper we stress the importance of a medium size (M-class) space mission, dubbed ``ASTROMEV", to fulfil these objectives.
\end{abstract}

\section{Introduction}

Gamma-ray astronomy has experienced a period of impressive scientific advances and successes during the last decade.  In the high-energy range studied with space instruments, above 100 MeV, the {\it AGILE} and {\it Fermi} missions led to important discoveries. In particular, the Large Area Telescope (LAT) of the {\it Fermi} satellite has established an inventory of over 5000  sources of various kinds (blazars, pulsars, supernova remnants, high-mass binaries, gamma-ray bursts (GRBs), etc.) showing a variety of gamma-ray emission processes \cite{4FGL}. Similarly, in the hard X-ray/low-energy gamma-ray band, the latest catalog of sources detected with the Burst Alert Telescope (BAT) of the {\it Neil Gehrels Swift} observatory contains 1632 sources in the range 14--195~keV \cite{Swift}. But at intermediate photon energies, between 0.2 MeV and 100 MeV, only a few tens of steady sources have been detected so far, mostly by the COMPTEL instrument on board the {\it Compton Gamma-Ray Observatory} (CGRO; see Ref.\cite{sch00}), such that this particular field of astronomy has remained largely unexplored.

Many of the most spectacular objects in the Universe have their peak emissivity at photon energies between 0.2 MeV and 100 MeV (e.g. gamma-ray bursts, blazars, pulsars, etc.), so it is in this energy band that essential physical properties of these objects can be studied most directly. This energy range is also known to feature spectral characteristics   associated with gamma-ray emission from pion decay, thus indicating hadronic acceleration. This fact makes the MeV energy region of paramount importance for the study of  radiating, nonthermal particles {\mt and for distinguishing leptonic from hadronic processes}. Moreover, this energy domain covers the crucial range of nuclear gamma-ray lines produced by radioactive decay, nuclear collision, positron annihilation, or neutron capture, which makes it as special for high-energy astronomy as optical spectroscopy is for phenomena related to atomic physics.

\begin{figure}[t!]
\centering
\includegraphics[width=0.8\textwidth]{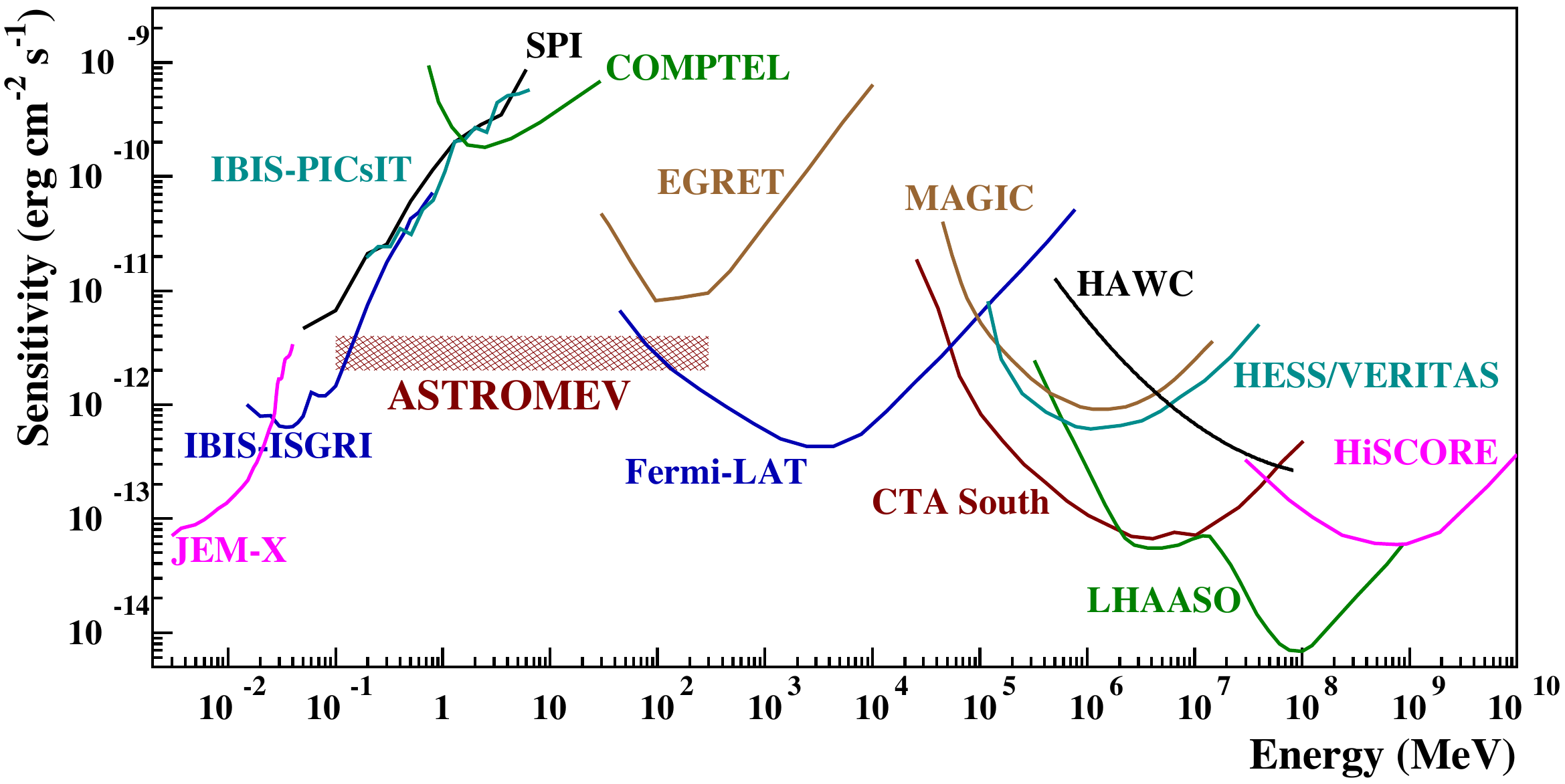}
\caption{Point source continuum differential sensitivity of different X- and $\gamma$-ray instruments (see \cite{science}). The hatched area indicates the targeted level of sensitivity of the next generation gamma-ray observatory for a source effective exposure of 1~year. 
\label{fig:sensi}}
\end{figure}

The fact that the MeV domain lags far behind compared to its neighbors (X-rays and gamma rays of high or very high energy) in terms of detection sensitivity (Fig.~\ref{fig:sensi}) is due to instrumental difficulties specific to this domain. In particular, below a few MeV, the lack of signature of the creation of an electron-positron pair is a limitation   to the capability to separate  gamma rays from charged particles and to evaluate the incoming direction of the photons. Moreover, while this is the domain of nuclear gamma-ray lines, which makes it extremely interesting for astrophysics, it also gives a strong instrumental background due to the deactivation of irradiated materials in space. 

However, recent progress in silicon detectors and readout microelectronics can allow the development of a new space instrument reaching a gain in sensitivity of about two orders of magnitude compared to CGRO/COMPTEL \cite{eastrogam}. In addition, the instrument can achieve excellent spectral and spatial resolution by measuring the energy and three-dimensional (3D) position of each interaction with the detectors. Such a mission, dubbed ``ASTROMEV'' in this White Paper, has the potential to answer key questions in astrophysics through a dedicated core science program:

\begin{itemize} \itemsep 0cm \topsep 0cm
\item \textbf{\em{Processes at the heart of the extreme Universe in the era of multi-messenger astronomy}}

Observations of relativistic jet and outflow sources (both in our Galaxy and
in active galactic nuclei, AGNs) in the X-ray and  GeV--TeV energy ranges have shown that the  MeV--GeV band holds the key to understanding the  transition from the low-energy continuum to a spectral range shaped by very poorly understood particle acceleration processes. 
{
\eam will:
(1) determine the composition (hadronic or leptonic) of the outflows and jets, which strongly influences the environment -- breakthrough polarimetric capability and spectroscopy 
providing the keys to unlocking this long-standing question;
(2) identify the physical acceleration processes in these outflows and jets (e.g. diffusive shocks, magnetic field reconnection, plasma effects), that may lead to dramatically different particle energy distributions; 
(3) clarify the role of the magnetic field in powering ultrarelativistic {jets in gamma-ray bursts}, through time-resolved polarimetry and spectroscopy.

In addition, measurements in the \eam energy band between 100 keV and 1 GeV will have a big  impact on multi-messenger astronomy in the 2030s. In particular, MeV energies are expected to be the characteristic cutoffs in Neutron Star - Neutron Star (NS-NS) and Black Hole - Neutron Star (BH-NS)  mergers, giving a decisive input to the study of the energetics of these processes. Moreover, a detector sensitive in the MeV region will allow the detection of the $\pi^0$ peak, disentangling hadronic acceleration mechanisms from leptonic mechanisms, and thus providing an independent input to neutrino astronomy.

\item \textit{\textbf{The origin and impact of high-energy cosmic-ray particles on Galaxy evolution}}

{ \eam   will  resolve the outstanding issue of the origin and propagation of low-energy cosmic rays affecting star formation. 
It  will measure cosmic-ray diffusion in interstellar clouds
and their impact on gas dynamics and state; it will provide crucial diagnostics about the wind outflows and their feedback on the Galactic 
environment (e.g., Fermi bubbles \cite{Su:2010qj}, Cygnus cocoon \cite{AckermannSB11}).}
\eam will have  optimal sensitivity and energy resolution to detect  line
emissions from 511 keV up to 10 MeV, and  a variety of 
issues will be resolved, in particular: (1)  origin
of the gamma-ray and positron excesses toward the Galactic
 inner regions;  (2)  determination of the astrophysical 
 sources of the local positron population from a very sensitive observation of pulsars and supernova remnants (SNRs). 
 As a consequence \eam will be able to discriminate the backgrounds to dark matter (DM) signals.

\item \textbf{\textit{Nucleosynthesis and the chemical enrichment
of our Galaxy}}

The \eam line sensitivity is more than an order of magnitude  better than previous instruments.  The deep exposure of the Galactic plane region will determine how different  isotopes are created in stars and distributed in the interstellar medium; it will also unveil the recent history of supernova explosions in the Milky Way.  
Furthermore, \eam will detect a significant number of Galactic novae and supernovae in nearby galaxies, thus addressing fundamental issues in the explosion mechanisms of both core-collapse and thermonuclear supernovae. The $\gamma$-ray data will provide a much better understanding of Type Ia supernovae and their evolution with look-back time and metallicity, which is a pre-requisite for their use as standard candles for precision cosmology.  

}
\end{itemize}

\vskip 2mm
\begin{table}
\begin{center}
\caption{\small Required instrument performance to achieve the core science objectives.
\label{tab:requirements}}
\begin{tabular}{| l | l |}\hline 
\textbf{Parameter} & \textbf{Value} \\ \hline \hline
Spectral range & 100~keV -- 1 GeV \\ \hline
Field of view & $\geq 2.5$ sr \\ \hline
Continuum flux sensitivity  & $<2 \times 10^{-5}$ MeV cm$^{-2}$ s$^{-1}$ at 1 MeV (any source) \\
for $10^6$ s observation time & $<5 \times 10^{-5}$ MeV cm$^{-2}$ s$^{-1}$ at 10 MeV (high-latitude source) \\
($3\sigma$ confidence level) & $<3 \times 10^{-6}$ MeV cm$^{-2}$ s$^{-1}$ at 500 MeV (high-latitude source) \\ \hline
Line flux sensitivity  & $<5 \times 10^{-6}$ ph cm$^{-2}$ s$^{-1}$ for the 511 keV line \\
for $10^6$ s observation time & $<5 \times 10^{-6}$ ph cm$^{-2}$ s$^{-1}$ for the 847 keV SN~Ia line \\
($3\sigma$ confidence level) & $<3 \times 10^{-6}$ ph cm$^{-2}$ s$^{-1}$ for the 4.44~MeV line from LECRs \\ \hline
 & $\leq 1.5^\circ$ at 1 MeV (FWHM of the angular resolution measure) \\
Angular resolution & $\leq 1.5^\circ$ at 100 MeV (68\% containment radius)  \\ 
 & $\leq 0.2^\circ$ at 1 GeV (68\% containment radius)  \\ \hline
Polarization sensitivity & Minimum Detectable Polarization $< 20$\% (99\% confidence level) \\
 & for a 10 mCrab source in $T_{\rm obs}=10^6$ s ($\Delta E=0.1 - 2$ MeV) \\ \hline
Spectral resolution & $\Delta E / E =3$\% at 1 MeV \\
 & $\Delta E / E =30$\% at 100 MeV \\ \hline
Time tagging accuracy & 1 $\mu$s (at $3\sigma$) \\
\hline
\end{tabular}
\end{center}
\end{table}

\section{Science case} \label{sec:science}

The photon energy range from 100 keV to 1 GeV is crucial in different sectors of astrophysics
(for a review, see \cite{science}). 


\subsection{The extreme extragalactic Universe}
\label{sec:extreme_uni}

The Universe contains objects with extreme properties that can be studied by measuring emission from particles that are accelerated near them. The emission is very intense, permitting measurements at very large distance, or redshift, when the Universe was young and many galaxies still forming. In many cases, a substantial fraction of the radiated power appears in the MeV band, and so a new gamma-ray observatory sensitive in the energy domain would offer an ideal view of the violent processes operating close by supermassive BHs, inside the powerful explosions that we see as gamma-ray bursts, and during the merger of binary neutron stars (NS). By deciphering many aspects of particle acceleration in the Universe, we address the question of why the energy distribution is so unbalanced: only a few particles carry an extreme share of the available energy, and by their feedback they shape numerous cosmic objects.

GRBs are explosive events with peak emission in the MeV band. Accurate polarimetry in this energy domain would permit measuring the structure and amplitude of the magnetic field that shapes the acceleration and transport of particles \cite{tat18}. Lorentz-invariance violation can be searched for \cite{grbliv}, and together with future gravitational wave detectors the relation between GRBs and the mergers of compact objects can be determined.

Clusters of galaxies are the largest gravitationally-bound structures in the Universe. In fact, they are still forming, leading to particle acceleration at structure formation shocks. Measuring their emission in the MeV band in conjunction with radio-band data lifts degeneracies in the interpretation and permits a precise study of the energy redistribution into magnetic field and accelerated particles, together with the feedback they impose on the cluster structure \cite{Brunetti14}.

\begin{figure}[t]
\centering
\includegraphics[width=0.7\textwidth]{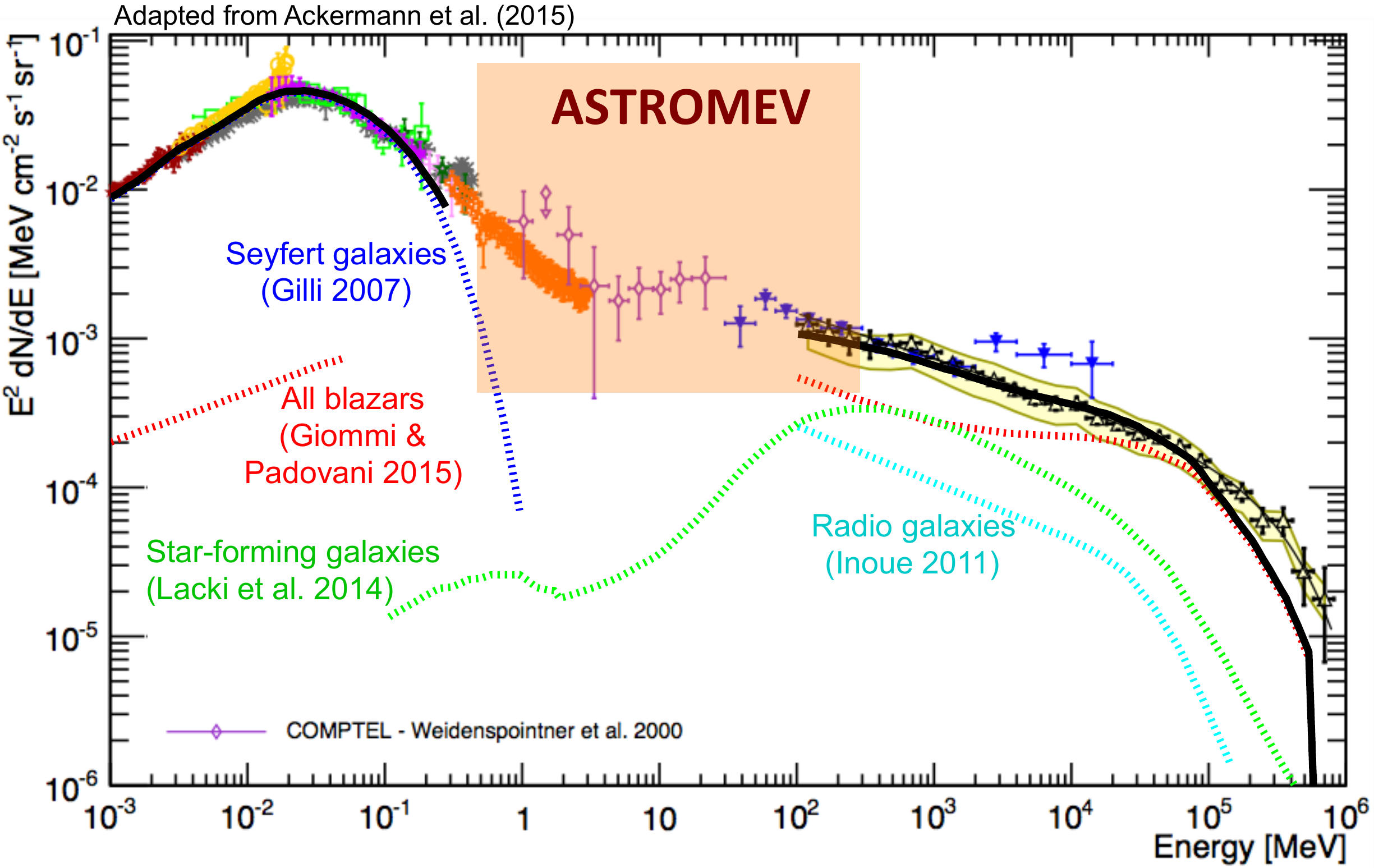}
\caption{Compilation of the measurements of the total extragalactic gamma-ray intensity between 1 keV and 820 GeV \cite{FermiEGB}, {with different components from current models}; the contribution from MeV blazars is largely unknown. The semi-transparent band indicates the energy region in which a new gamma-ray observatory could dramatically improve on present knowledge.\label{fig:egb}}
\end{figure}

The MeV gamma-ray background (see Fig.~\ref{fig:egb}) contains invaluable collective information about nucleosynthesis in distant SNe, DM annihilation, 
and supermassive BHs. The latter are often visible also visible as AGN, and they are the most luminous persistent sources in the Universe, many of which emit their bulk power in the MeV band. A sensitive observatory in this energy domain can use these unique beacons to study the formation history and evolution of supermassive BHs at times when the Universe had only a fraction of its current age. MeV-band observations address the energy limit to which electrons may be accelerated, the location where this happens. By studying the spectral response to changes in the activity of these objects, we can distinguish the emission from electrons from that of energetic ions. The MeV band is ideally suited for this inquiry, because emission at higher gamma-ray energies may be absorbed
by Extragalactic Background Light (EBL), and the specific contribution from photo-pair-production by high-energy cosmic nuclei is a critical discriminant in the soft gamma-ray band, as an analysis of the recent detection of a statistical association of a 300-TeV neutrino event \cite{txsicecube} with an extended gamma-ray flare of the Active Galactic Nucleus TXS0506+056 shows \cite{txsmulti}. Finally, the MeV band carries the cascade emission of all the absorbed Very-High-Energy (VHE) gamma-ray emission that is emitted in the Universe, and so its study provides a unique view of its extreme particle acceleration history, including the feedback on the intergalactic medium and the magnetic-field genesis therein.

Last but not least, the MeV range is the perfect companion for multi-messenger astronomy.  On top of the Spectral Energy Distribution (SED) of the electromagnetic (EM) emission  by TXS0506+056, mentioned before, the recent NS-NS merger generating the GW170817 event and the corresponding gamma-ray signal detected by \textit{Fermi} GBM and \textit{INTEGRAL} has shown that the EM cutoff of this class of mergers is in the MeV range \cite{multins}. Furthermore, for a sufficiently close  event, ASTROMEV could detect the continuum and nuclear line emissions expected from a kilonova (KN) following a merger event like GW170817. This remarkable event highlighted the importance of KNs in the nucleosynthesis of heavy elements by the r process \cite{pia17,abb17}. The predicted line emission in the MeV band \cite{hot16,li19} could be detected with ASTROMEV up to a maximum distance of $\sim 10$~Mpc. The expected rate of KNs is still quite uncertain: $200$--$400$ KN Gpc$^{-3}$ yr$^{-1}$ if associated with GRB emission \cite{del18} and $\sim 1500$ KN Gpc$^{-3}$ yr$^{-1}$ as Gravitational Wave (GW) detection only \cite{abb17}. However, if the prompt emission of the KN is associated with substantial gamma-ray emission, absorption edges, possibly variable in time, from freshly-formed elements from the immediate environment of the source may also be detected (e.g., \cite{ama00}). 


ASTROMEV could allow to detect a sub-class of short GRBs having a peculiar origin, i.e., a transition from a Neutron star (NS) to a more compact stellar object called a Quark Star (QS) \cite{witten,perez}. Distinctive signatures compared to the binary (NSNS or NSBH) merger scenario are the shortness of the prompt gamma-ray emission, estimated to be $\sim 0.1$ s and peak energies in excess of $\sim 100$ keV with possibly a thermal spectrum and spectral features due to the heavy composition of ejecta. In addition the associated GW emission would be  different from the quoted binary merger (NS-NS, NS-BH) events. 

\subsection{Cosmic ray interactions}
\label{sec:cosmic_ray}

\begin{figure}[t]
\centering
\includegraphics[width=0.8\textwidth]{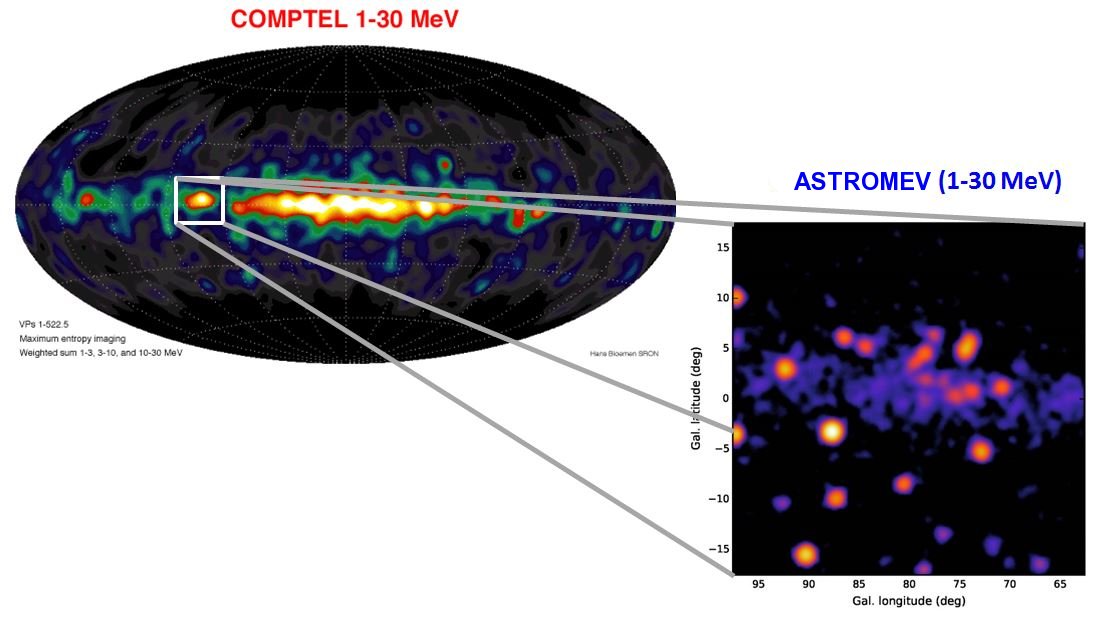}
\includegraphics[width=0.9\textwidth]{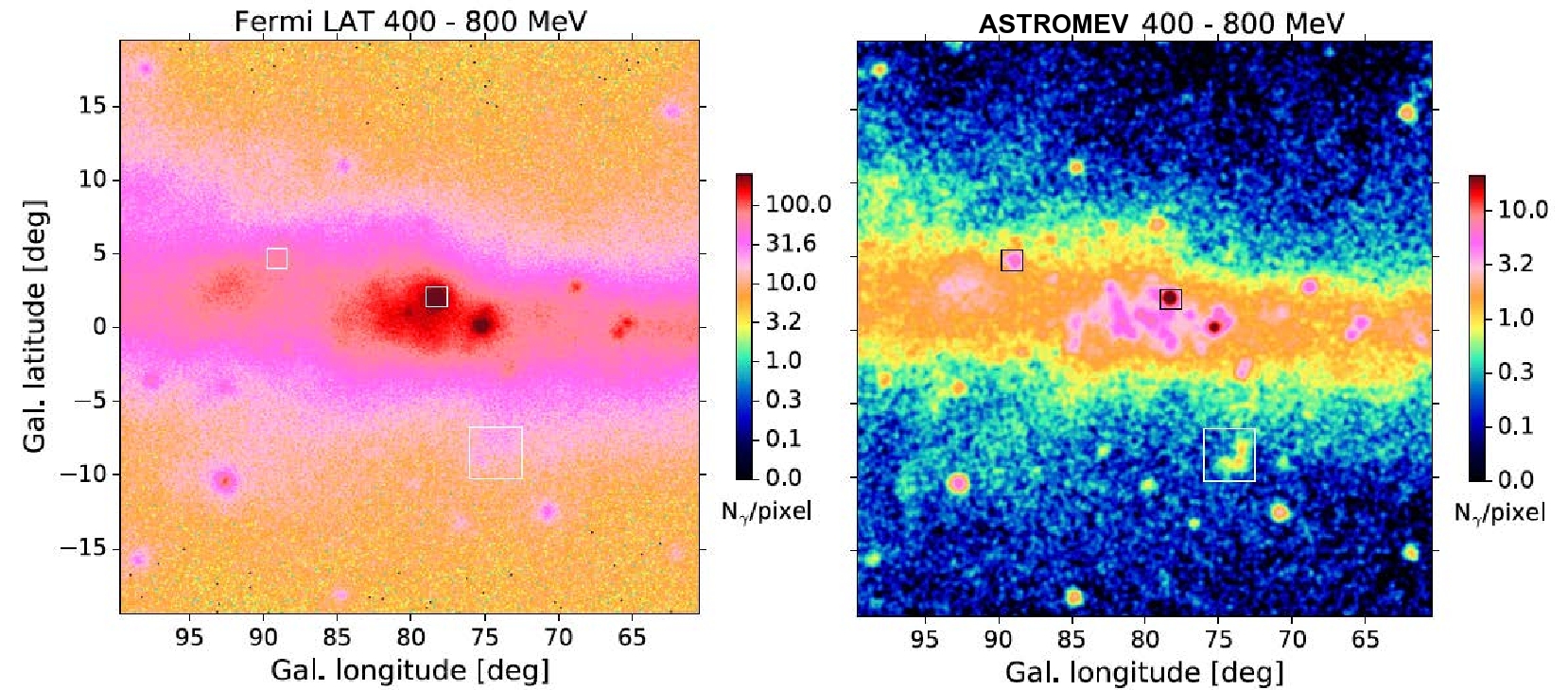}
\caption{An example of the capability of \eam to transform our 
 knowledge of the MeV-GeV sky. Upper panel: The upper left figure shows the
1-30 MeV sky as observed by COMPTEL in the 1990s; the upper right figure
  shows the simulated Cygnus region in the 1-30 MeV
energy region  from \eamp. Lower panel: comparison between the view of the Cygnus region by \textit{Fermi} in 8 years (left) and that by ASTROMEV in one year of effective exposure (right) between 400 MeV and 800 MeV.}
\label{fig:cygnus}
\end{figure}

A clear understanding of the origin and evolution of cosmic rays (CRs) is still missing despite one century of impressive observational discoveries and theoretical progress \cite{bla13}. Understanding their origin is an interdisciplinary problem involving fundamental plasma physics, to describe the diffusive shock acceleration process, as well as astrophysical and particle-physics diagnostics, to characterize the particle properties and the local conditions in the acceleration zones. While we still lack a reliable explanation for the existence of CRs near and beyond PeV energies in the Milky Way and beyond EeV energies in  extragalactic space, we also  know very little about the Galactic population of low-energy CRs, with energies below a few GeV per nucleon. We still need information on their sources and injection spectra into the interstellar medium, on their transport properties and flux distribution at all interstellar scales in the Galaxy, and on their impact on the overall evolution of the interstellar medium and on the dynamics of Galactic outflows and winds. The performance of a new mission such as \eam would provide unique results in a number of important CR issues. 

Sensitive observations of a set of CR sources, such as young SNRs, across the MeV--GeV bandwidth, would allow for the first time to distinguish the emission produced by the interactions of CR nuclei with the ambient gas and the non-thermal emission from CR electrons \cite{giuliani11_W44,ackermann13_W44,cardillo14_W44,jogler16,2011NatCo...2E.194M}. Combined with high-resolution radio and X-ray observations of the remnants (e.g., \cite{2008ARA&A..46...89R}),  gamma-ray data in the MeV--GeV region would provide information on CR injection into the acceleration process, on the structure of magnetic fields inside the remnants, and on the spectrum of CRs freshly released into surrounding clouds.

\textit{Fermi} LAT could resolve only one case of CR activity in a Galactic super-bubble to study the collective effects of multiple supernovae and powerful winds of young massive stars \cite{AckermannSB11}. An improved angular resolution in the sub-GeV range would provide more  case studies (see Fig.~\ref{fig:cygnus} for an illustration with the Cygnus region), individually as well as collectively in the inner Galaxy, which would help to probe the interplay between CRs and the turbulent medium of star-forming regions during the early steps of their Galactic voyage \cite{Bykov2014,Grenier15}. Individual massive binary stars like $\eta$ Carinae, which is one of the most luminous massive binary systems in the Galaxy and the likely progenitor of the next Galactic supernova, are promising candidates to study particle acceleration by their powerful winds \cite{2017arXiv170502706B}. Following their time variability from radio to gamma-ray energies can provide key diagnostics on the acceleration efficiency.

The \textit{Fermi} Bubbles are one of the most spectacular and unexpected discoveries based on the \textit{Fermi}-LAT data \cite{Su:2010qj,Fermi-LAT:2014sfa}. However, the origin of these gigantic lobes above and below the Galactic Center (GC) is still unknown: possible sources are  outflows from the supermassive black hole Sgr~A* (AGN scenario) or combined wind from massive star activities and supernova explosions in the central molecular zone (starburst scenario). Improving the angular resolution relative to the \textit{Fermi}-LAT PSF will be essential in the derivation of the shape of the {\em Fermi} Bubbles at energies below 1 GeV, and to detect the expected spectral differences between the AGN (leptonic) and starburst (hadronic) models. 

CR nuclei of energies below a few GeV per nucleon contain the bulk energy density of the Galactic CRs. They are the main source of ionization and heating in the highly obscured star-forming clouds that are well screened from UV radiation. At the same time they are the source of free energy and pressure gradients to support large-scale magnetohydrodynamic (MHD) outflows and Galactic winds that control the overall evolution of a galaxy \cite{Grenier15,Pakmor16}. MeV gamma-ray observations of the inner Galaxy with the targeted sensitivity (Fig.~\ref{fig:sensi}) would provide the first nuclear spectroscopic data on the low-energy CR population \cite{ben13}. The energy coverage of the telescope would also allow a precise separation of the CR nuclei and electron/positron populations (and spectra) across the Galaxy. The higher-resolution images (see Fig.~\ref{fig:cygnus}) would shed light on the degree of correlation between the CR distributions and stellar activity, at the scale of cloud complexes up to that of spiral arms, in order to better constrain the diffusion properties of CRs in a galaxy (e.g., \cite{Grenier15,nava17}). 

Last, but not least, maps of the total interstellar gas mass inferred from CRs and the GeV data from a new mission with a targeted resolution $<12'$ at 1 GeV (see Table~\ref{tab:requirements}) would serve a broad  community wishing to improve the calibration of gas tracers (radio and dust tracers) in a large variety of cloud states \cite{Planck15}.

\subsection{Explosive nucleosynthesis and chemical evolution of the Galaxy}
\label{sec:nucleo}

Exploding stars play a very important role in astrophysics since they inject important amounts of kinetic energy and newly synthesized chemical elements into the interstellar medium in such a way that they completely shape the chemical evolution of galaxies. Furthermore, the ``pyrotechnical'' effects associated with such outbursts can be so bright and regular that they can be used to measure distances at the cosmological scale. For instance, Type Ia SNe (SNIa) led to the discovery that the Universe was expanding in an accelerated way \cite{rie98,per99}.  

The majority of  outbursts are associated  with instabilities of electron degenerate structures in single stars (core collapse and electron capture supernovae) or when they accrete matter from a companion in a close binary system (SNIa and classical novae, for instance). Systematic research on transient events  have revealed a surprising variety of outbursts that goes from ``Ca-rich'' transients, placed in the gap between Type Ia SNe and novae, Type Iax, ``02es-like'' SNe, ``super-Chandrasekhar'' SNe in the domain of the so-called thermonuclear SNe \cite{hil00,hi13}, to, e.g., Type IIn, Type In, and so-called ``impostors'' in the domain of core collapse of massive stars \cite{woo05,jan12,bur13}.

Many of these events, if not all, imply the activation of thermonuclear burning shells that synthesize new isotopes, some of them radioactive. As the ejecta  expand, more and more photons avoid thermalization and escape, such that they can be used as a diagnostic tool. Each  of the different explosion scenarios leads to differences in the intrinsic properties of the ejecta, like the density and velocity profiles, and the nature and distribution of the radioactive material synthesized. This translates into differences in the light curves and line widths of the expected gamma-ray emission. Therefore, the observation with gamma rays becomes a privileged diagnostic tool with respect to other measurements thanks to the penetration power of high energy photons and the association of gamma-ray lines to specific isotopes created by the explosion \cite{isern16}.  

\begin{table}[t]
\begin{center}
\caption{\small Star-produced radioisotopes relevant to gamma-ray line astronomy.
\label{tab:radioisotopes}}
\begin{tabular}{|c|c|c|c|c|}
\hline
{\bf Isotope} & {\bf Prod. site$^{\rm a}$} & {\bf Decay chain$^{\rm b}$} & {\bf Half-life$^{\rm c}$} &
{\bf $\gamma$ ray energy (keV)} \\
  & & &
  & {\bf and intensity$^{\rm d}$} \\\hline  \hline
$^7$Be & Nova & $^7$Be~$\stackrel{\epsilon}{\longrightarrow}$~$^7$Li*
& 53.2~d & 478~(0.10) \\\hline
$^{56}$Ni & SNIa, CCSN &
$^{56}$Ni~$\stackrel{\epsilon}{\longrightarrow}$~$^{56}$Co*
& 6.075~d & 158~(0.99), 812~(0.86) \\
& & $^{56}$Co~$\stackrel{\epsilon(0.81)}{\longrightarrow}$~$^{56}$Fe*
& 77.2~d & {847}~(1), {1238}~(0.66) \\\hline
$^{57}$Ni & SNIa, CCSN &
$^{57}$Ni~$\stackrel{\epsilon(0.56)}{\longrightarrow}$~$^{57}$Co*
& 1.48~d & 1378~(0.82) \\
& & $^{57}$Co~$\stackrel{\epsilon}{\longrightarrow}$~$^{57}$Fe*
& 272~d & {122}~(0.86), {136}~(0.11) \\\hline
$^{22}$Na & Nova &
$^{22}$Na~$\stackrel{\beta^+(0.90)}{\longrightarrow}$~$^{22}$Ne*
& 2.60~y & {1275}~(1) \\\hline
$^{44}$Ti & CCSN, SNIa &
$^{44}$Ti~$\stackrel{\epsilon}{\longrightarrow}$~$^{44}$Sc*
& 60.0~y & {68}~(0.93), {78}~(0.96) \\
& & $^{44}$Sc~$\stackrel{\beta^+(0.94)}{\longrightarrow}$~$^{44}$Ca*
& 3.97~h & {1157}~(1) \\\hline
$^{26}$Al & CCSN, WR &
$^{26}$Al~$\stackrel{\beta^+(0.82)}{\longrightarrow}$~$^{26}$Mg*
& 7.2$\cdot$10$^5$~y & {1809}~(1) \\
 & AGB, Nova & & & \\\hline
$^{60}$Fe & CCSN &
$^{60}$Fe~$\stackrel{\beta^-}{\longrightarrow}$~$^{60}$Co*
& 2.6$\cdot$10$^6$~y & 59~(0.02) \\
& & $^{60}$Co~$\stackrel{\beta^-}{\longrightarrow}$~$^{60}$Ni*
& 5.27~y & 1173~(1), 1332~(1) \\\hline
\end{tabular}
\begin{minipage}{0.92\linewidth}
{\vspace{1 mm} \small
{$^{\rm a}$ Sites which are believed to produce observable
gamma-ray line emission. Nova: classical nova; SNIa: thermonuclear
SN (type Ia); CCSN: core-collapse SN; WR: Wolf-Rayet star;
AGB: asymptotic giant branch star.}

{$^{\rm b}$ $\epsilon$: orbital electron capture. When an
isotope decays by a combination of $\epsilon$ and $\beta^+$ emission, only 
the most probable decay mode is given, with the corresponding fraction in 
parenthesis.}

{$^{\rm c}$ Half-lives of the isotopes decaying by 
$\epsilon$ are for the neutral atoms.}

{$^{\rm d}$ The values in brackets correspond to the number of photons emitted in the gamma-ray
line per radioactive decay.}
}
\end{minipage}
\end{center}
\end{table}

\begin{figure}
\centering
\includegraphics[width=0.6\textwidth]{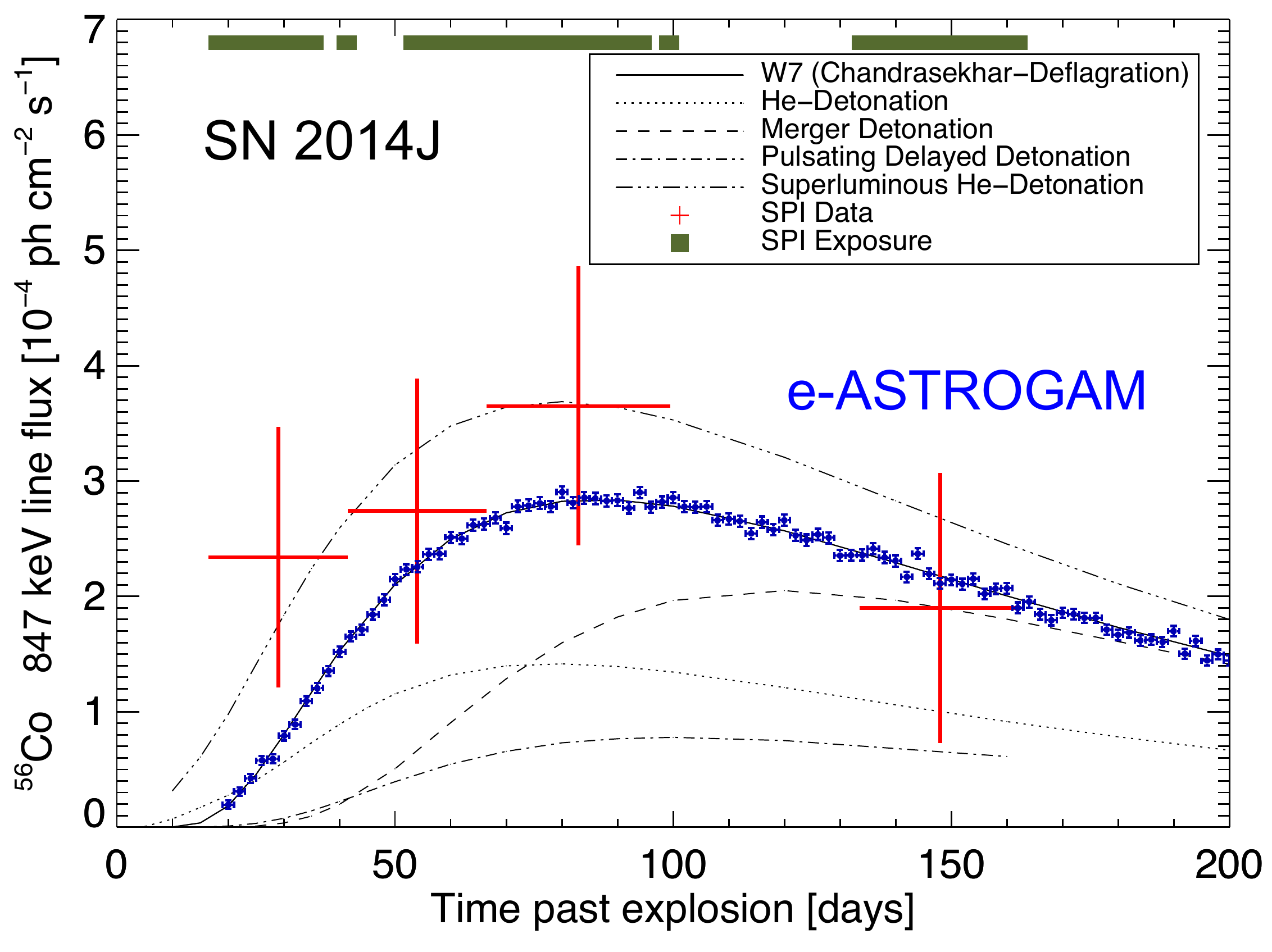}
\caption{\small Light curve of the 847 keV line from $^{56}$Co decay in SN~2014J. \textit{INTEGRAL} data (adapted from Fig.~4 in Ref.~\cite{diehl15}, red data points) are compared to various models of Type Ia SN \cite{the14}. A simulation of the e-ASTROGAM response \cite{eastrogam} to a time evolution of the 847 keV line such as in the W7 model \cite{nomoto84} shows that the sensitivity improvement by a new gamma-ray space mission (blue points) can lead to a much better understanding of the SN progenitor system and explosion mechanism.}
\label{fig:SNIa}
\end{figure}

Table~\ref{tab:radioisotopes} displays the main detectable gamma-ray line emissions expected in several nucleosynthesis events (see Ref.~\cite{die11} and references therein). The radio-isotopes with a relatively short lifetime can be used to directly characterize the individual explosion events or the first stages of the remnant, while the long-lived ones, i.e., with lifetimes much longer than the characteristic time between events, will produce a diffuse emission resulting from the superposition of many sources that can provide information on stellar nucleosynthesis \cite{die06,die16}, but also on the physical conditions and dynamics of the Galactic interstellar medium (see, e.g., \cite{kra15}).

\begin{figure}[t]
\begin{center}
\includegraphics[width=0.6\textwidth]{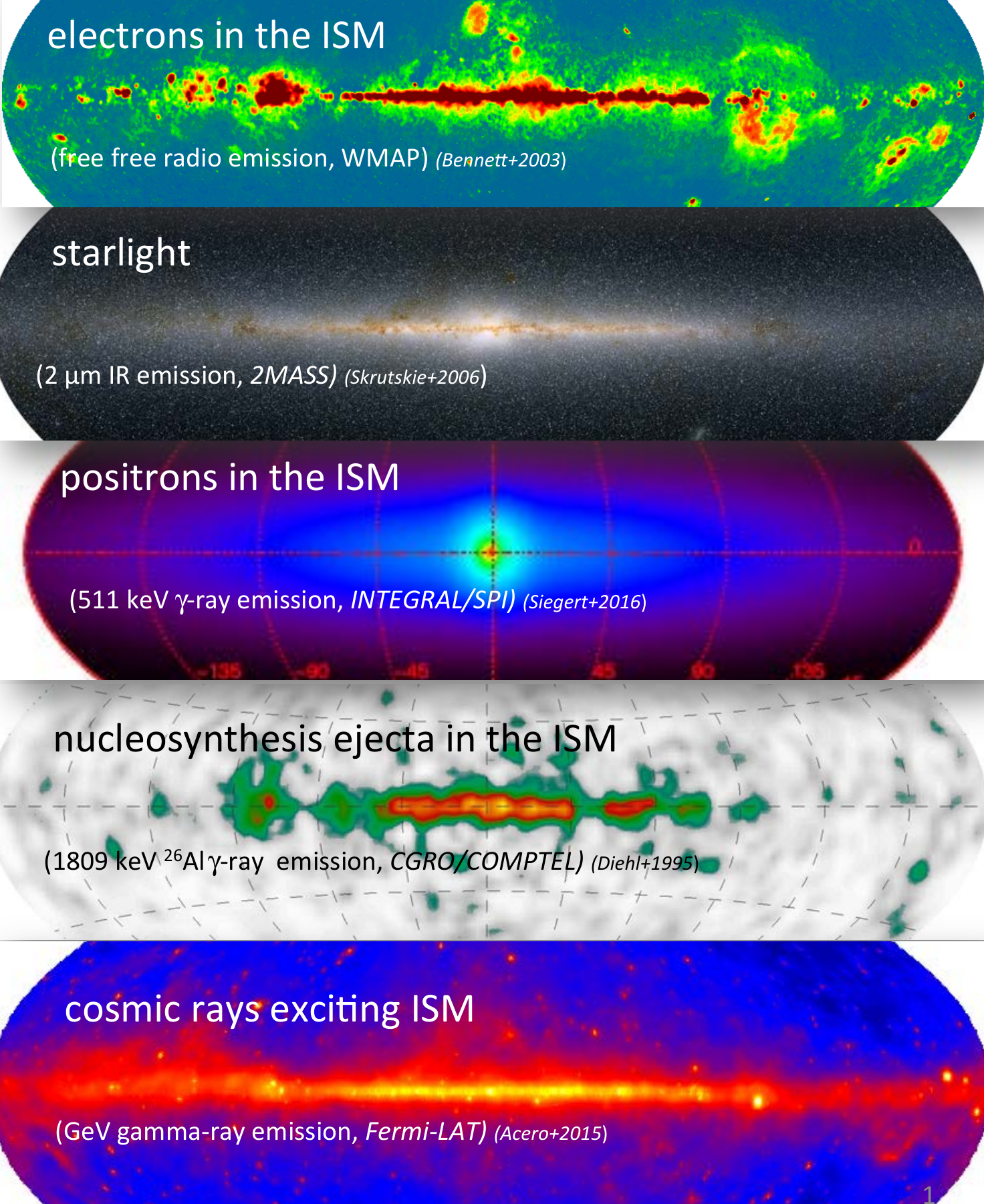}
	\caption{\small The diffuse emissions of our Galaxy across several astronomical bands: ASTROMEV will explore the link between starlight (second image from top) and CRs (top and bottom). The current-best images of positron annihilation (3rd from top) and $^{26}$Al radioactivity (4th from top) gamma rays illustrate that this link is not straightforward, and e-ASTROGAM will uncover more detail about the astrophysical links and processes. (Image composed by R. Diehl, from observations with WMAP, 2MASS, \textit{INTEGRAL}, CGRO, and \textit{Fermi}; Refs. \cite{3FGL,Bennett2003,Skrutskie2006,sie16,Diehl95})}
\label{fig:diffuse}
\end{center}
\end{figure}

It is important to distinguish here between guaranteed and opportunity observations. By guaranteed, we understand observations that can be predicted with enough anticipation and with the certitude that they can be included into the ordinary mission scheduling. Three examples of guaranteed observations would be:
\begin{enumerate}
\item Measurement of the total mass of $^{56}$Ni/$^{56}$Co ejected by SNIa. This value is fundamental to calibrate the Phillips \cite{phi93} relation and the yield of synthesized Fe. The explosion time and location are not known a priori, but thanks to the sensitivity of a new gamma-ray observatory \cite{eastrogam}, it is expected that about a dozen SNIa will occur at a distance smaller than 35 Mpc in three years of mission. The observations will have to be performed around 50--100 days after the explosion, when all the SN properties (subtype, luminosity,...) will already be known \cite{chur14,chur15,diehl15}. ASTROMEV will achieve a major gain in sensitivity compared to \textit{INTEGRAL} for the main gamma-ray lines arising from $^{56}$Ni and $^{56}$Co decays (Fig.~\ref{fig:SNIa}) allowing, for events like SN~2014J, the exquisitely accurate (at percent level) measurements of the  Ni mass, the mass of the progenitor and the expansion velocity, easily differentiating between major astrophysical scenarios. 
\item Clumping degree of core-collapse SNRs as a diagnostic of internal asymmetries \cite{mah88,tue90}. This property can be obtained from the radioactive emission of the $^{44}$Ti/$^{44}$Sc chain \cite{gre14,gre17}. The targeted sensitivity would allow the detection of this emission in all young Galactic SNRs and in the remnant of SN1987A.
\item Mapping of the positron annihilation radiation \cite{joh72,lev78,kno05,wei08,sie16} and the long-lived isotopes $^{26}$Al and $^{60}$Fe \cite{Diehl06,Wang07,Wang09,mar09,Diehl13}. The expected huge increase in sensitivity compared to current gamma-ray missions should allow the building of detailed maps of these Galactic diffuse emissions (see Fig.~\ref{fig:diffuse}), which will shed a new light on nucleosynthesis in massive stars, SNe and novae, as well as on the structure and dynamics of the Galaxy \cite{kra15}. Individual objects (e.g., SNRs) should also be detected in these lines.
\end{enumerate} 
Given the explosive nature of the events considered here, the majority of the observations will belong to the category of Targets of Opportunity (ToO). The information and the relevance of the observation will depend on the distance of the events. Two examples would be:
\begin{enumerate}
\item Novae. The targeted sensitivity would allow the detection of the $^{22}$Na (1275 keV) line to a distance large enough to observe about one nova per year, but that of the $^{7}$Be (478 keV) line demands a shorter distance and is thus uncertain during the three years of nominal mission duration. Therefore the results that can be obtained from every individual event will depend not only on the nature of the event, but also on the distance \cite{CH74,Gom04,Her08}.
\item Type Ia and Core-collapse SNe. The detection of the early gamma-ray emission before the maximum optical light in the SNIa case \cite{diehl14,isern16} and the determination of the amount of $^{56}$Ni ejected by CCSN \cite{mat88} would be fundamental to understanding the nature of the progenitor in the first case and of the explosion mechanism in both cases. Given the expected sensitivity, it is foreseen to detect these details to a distance of about ten Mpc, which ensures the detection of several events and opens the possibility of comparing SN subtypes. 
\end{enumerate}
The observation of ToOs is unpredictable, but extremely rewarding if successful, and exploding stars and related phenomena are within this category. It is important to realize that the targeted increase of sensitivity would guarantee that a 
significant number of events will be observed in an effective way.

\subsection{Observatory science in the MeV domain}

Since the MeV domain is largely unexplored, the observatory science could be particularly interesting. We summarize here some of the topics, referring the reader  to \cite{science} for a more complete treatment.

\subsubsection{Physics of compact objects}
\label{sec:comp_obj}

Neutron stars (NSs) and black holes (BHs) are the most compact objects in the Universe, capable  of distorting the structure of the space-time around them. They are observed to manifest themselves in a great variety of ways: pulsating and bursting, accreting from a binary companion, interacting with its wind, or even merging with it. NSs are found both in binary systems, often with other compact stars such as white dwarfs or NSs, or as isolated sources. The most extreme neutron stars, the explosive magnetars, are found with the same outward characteristics, such as spin period and in some cases also surface magnetic field, as those of more placid rotation-powered pulsars, but they show a spectacular bursting and flaring activity in the gamma-ray band. Understanding the evolutionary link between different NSs classes and their inter-relation is one of the holy grails of compact object astrophysics. Observations at MeV energies can uniquely address this issue. 

The magnetar-like phenomenology is likely caused by a twisted toroidal magnetic field structure capable of releasing a power larger than that of dipole spin-down and of causing instabilities, magnetic field reconnection, and crustal fractures that ultimately result in their spectacular flaring emission. Many magnetars have hard non-thermal components extending to at least 100~keV with no observed cutoffs, although one is expected in the MeV band from COMPTEL upper limits. Particle acceleration along closed field lines and its highly non-linear competition with attenuation from photon splitting and pair production are not yet understood, nor is the geometry of the magnetic lines bundle where currents flow. The exploration of such hard tails and cutoffs as well as their phase-resolved behavior and polarization with a sensitive, large throughput gamma-ray space mission is key to resolving these issues. 

Another NS puzzle is the nature of pulsar gamma-ray emission. Fermi revolutionized gamma-ray pulsar studies increasing the number of pulsars detected above 100~MeV from 7 with CGRO/EGRET to about 200 today \cite{Fermipulsars}. However, in the soft gamma-ray region there are only 18 detections above 20~keV and only four have been detected with pulsed emission in the range 1~--~10~MeV.  Such MeV pulsars appear to have the peaks of their spectral energy distributions at MeV energies, so the clues to their nature lie in measurements by more sensitive detectors like ASTROMEV. Thus, as Fermi did at higher energies, ASTROMEV can  revolutionize the number of detected sources and our understanding of pulsar physics at this energy. 

The application of pulsar emission models to current data is plagued by the poor knowledge of pulsar inclination and viewing geometry for most sources. The expected polarization signature, in fact, depends significantly on the geo\-metry of the system and the location of the emitting zones. Gamma-ray polarization measurements with ASTROMEV will be crucial to nail down the system inclination (magnetic and spin axis with respect to the proper motion), reveal the magnetic field topology, locate the emission region(s) in the magneto\-sphere and identify the emission mechanism. Particularly important will be ASTROMEV information on the misalignment between spin and proper motion axis, which is still highly debated and is linked to the way in which the kick is imparted to a proto-neutron star during its formation and to the duration of the physics of the acceleration phase (see, e.g., \cite{Spruit98,Noutsos12,Mignani17}). Another key information is the misalignment between spin and magnetic field axis, which is crucial to quantify the contribution of pulsars to GW emission since orthogonal rotators will be efficient sources of gravitational radiation  \cite{Cutler02,Stella05,DallOsso09,Lasky15}. 

A number of pulsars in binary systems are thought to have intra-binary shocks between the pulsar and companion star that can accelerate particles of the pulsar wind to greater than TeV energies.  Gamma-ray binaries, with a young rotation-powered pulsar in orbit around a massive Be star, show orbitally modulated emission at radio, X-ray, GeV, and TeV energies.  Models with either inverse-Compton or synchrotron radiation can fit the X-ray to GeV spectrum and better measurements at MeV energies would constrain the mechanism.  Observations of accreting X-ray binaries, that contain either NS-NS or NS-BHs, at MeV energies can uncover the emission mechanisms that are operating as well as the role of the jets in these sources.  An exciting possibility is the detection of a 2.2~MeV neutron-capture line coming from the inner parts of the accretion disc or from the NS atmosphere, which would be a major discovery and give new constraints on accretion physics and the gravitational redshift at the NS surface, respectively.

Binaries containing millisecond pulsars and low mass companions also show orbitally-modulated X-ray emission from intra-binary shocks and three of these are observed to transition between rotation-powered and accretion-powered states \cite{Papitto13,Stappers14,Roy15}. Observations with the proposed ASTROMEV observatory will fill in the spectral gap from 0.1~--~100 MeV to help us understand the nature of these transitions and the limits to acceleration in the pulsar wind shock. 

Many millisecond pulsars are found in globular clusters. {\em Fermi} has discovered both  gamma-ray emission from many clusters and also pulsations from pulsars within some clusters.  The nature of the diffuse X-ray and TeV emission detected from several clusters is presently a mystery and could come from magnetospheric emission or from electron-positron pairs ejected from the pulsars in the cluster. ASTROMEV will map the extent of the diffuse X-ray component in the MeV range, settling the crucial question of its origin which is still unanswered by current data.

\subsubsection{Solar- and Earth-science}
\label{sec:solar}

The same gamma-ray emission mechanisms at play in celestial sources can be studied in more detail, even if in different environmental conditions, in local gamma-ray sources such as those present in the Solar System. In particular the interactions of CRs with radiation fields and matter, at the Sun and with other Solar System  bodies, such as the Moon, the acceleration of particles and their emission in the upper atmosphere, the physics of magnetic reconnection and particle acceleration in solar flares are examples of science objectives that can be explored by observing gamma rays coming from the Sun, the Moon, the Earth, and other bodies in the Solar System.  

Solar flares are the most energetic phenomena in the Solar System. These events are often associated with explosive Coronal Mass Ejections (CMEs). The frequency of both flares and CMEs follows the 11-year solar activity cycle, the most intense ones usually occurring during the maximum. What triggers the flares is presently not completely understood. Flare energy may be considered to result from reconnecting magnetic fields in the corona. Phenomena similar to solar flares and CMEs are believed to occur at larger scales elsewhere in the Universe, such as in AGNs \cite{DiMatteo98}. These energetic phenomena from the Sun are therefore the most accessible laboratories for the study of the fundamental physics of transient energy release and efficient particle acceleration in cosmic magnetized plasmas. The gamma-ray emission from solar flares results from the acceleration of charged particles which then interact with the ambient solar matter in the regions near the magnetic field lines. Accelerated electrons mainly produce soft and hard X-rays via non-thermal bremsstrahlung.   Accelerated protons and ions emit at higher energies: nuclear interactions produce excited and radioactive nuclei, neutrons and pi-mesons. All of these products subsequently are responsible for the gamma-ray emission via secondary processes, consisting of nuclear gamma-ray lines in the 1-10 MeV range and a continuum spectrum above 100 MeV. The high-energy gamma-ray emission light curve can be similar to the one observed in X-rays, lasting for 10--100 s and indicating the acceleration of both ions and electrons from the same solar ambient. This is referred to as the ``impulsive" phase of the flare. However, some events have been found to have a long-duration gamma-ray emission, lasting for several hours after the impulsive phase.  
A new gamma-ray mission covering a very broad energy range, from about 100 keV to 3 GeV, will have the opportunity 
to study the evolution in time of the hard-X and gamma-radiation from each solar flare event, helping to constrain models of acceleration and propagation. It will 
detect the de-excitations lines from accelerated ions, which will be fundamental to gain insight into the chemical abundances and about the physical conditions where accelerated ions propagate and interact. Spectral analysis at higher energies will also allow disentangling the electron bremsstrahlung and pion-decay components. A polarized bremsstrahlung emission in hard X-ray from solar flares is expected if the phase-space distribution of the emitting electrons is anisotropic with important implications for particle acceleration models. 

The Moon is one of the brightest sources of high-energy gamma rays in the Solar System. Gamma rays from the Moon originate in the shower cascades produced by the interactions of Galactic CR nuclei with the lunar surface. The lunar gamma-ray emission depends on the fluxes of the primary cosmic-ray nuclei impinging on the Moon and on the mechanisms of their hadronic interactions with the rock composing the lunar surface. In addition to providing a new accurate measurement of the lunar gamma-ray spectrum in the sub-GeV band, the proposed observatory will extend the energy range observed by previous missions towards lower energies. This feature will provide the unique opportunity to explore possible gamma-ray lines in the hundreds of keV to MeV region, originating from the decays of excited states produced in the interactions of CR nuclei with the lunar rock. Measurements of the gamma-ray flux from the Moon also provide a useful tool to study the properties of CRs and to monitor the solar cycle, since it depends on the primary CR nuclei fluxes, which change with the solar activity.  The lunar gamma-ray data at low energies will also represent a powerful tool to monitor the solar modulation and to study the CR spectra impinging on the Moon surface. 

Terrestrial gamma-ray flashes (TGFs) are very intense gamma-ray emission episodes coming from the upper atmosphere and strongly correlated with lightning activity. They are generally interpreted as bremsstrahlung high-energy radiation emitted by free electrons in the air, accelerated to relativistic energies by intense electric fields presents in the atmosphere under thunderstorm conditions. The importance of gamma-ray observations from space satellites flying in Low Earth equatorial orbit is based on the possibility of detecting TGFs in the tropical regions where the frequency of thunderstorms is higher. 
Gamma-ray observations should also confirm the possible presence of a high-energy population of TGFs emitting at energies greater than 40 MeV. 

\subsubsection{Fundamental physics}
\label{sec:fund_phys}

The topic of fundamental physics in the context of high-energy astrophysics is often related to fundamental symmetries of nature which can be studied over cosmological distances, at high energies and in extreme environments. 

Gamma rays as a probe have been used for a variety of subjects in fundamental physics, the  most studied question for gamma-ray observations in general being the quest for dark matter (DM). The exploration of topics in fundamental physics that can be addressed with a new observatory in the gamma-ray MeV range is gaining momentum:
axion-like particles and primordial black-holes as well as possible observations elucidating the question of matter-antimatter asymmetry and, last but not least, different aspects of searches for DM particles with some focus on small masses.

The existence of DM is by now established beyond reasonable doubt, see e.g. \cite{bergstrom12,Ade:2015xua}, however its nature is one of the most pressing questions in science today.  Among the most popular DM candidates are weakly interacting massive particles (WIMPs), with masses and coupling strengths at the electroweak scale. Besides the fact that many of these are theoretically very well motivated, such as the supersymmetric neutralino \cite{Jungman:1995df}, an attractive feature of this class of candidates is that the observed DM abundance today can straightforwardly be explained by the thermal production of WIMPs in the early Universe. WIMPs are searched for by a variety of techniques:  directly by placing sensitive detectors in underground locations with the aim to detect WIMP-induced nuclear recoils and indirectly by detecting the secondary products of WIMP annihilation or decay.

WIMP candidates can also be produced at the Large Hadron Collider (LHC) by proton-proton collisions, which then would need to be confirmed by astrophysical observations. The latest LHC results, based on almost 40 fb$^ {-1}$ of data at $\sqrt{s} $= 13 TeV (e.g. \cite{Aaboud:2017yqz}) did not reveal any sign of WIMP DM. In indirect detection the \textit{Fermi} Large Area Telescope managed to push the sensitivity below the canonical thermal WIMP cross-section for WIMPs in the mass range from about 5 GeV to 100 GeV without firmly confirmed detection. There is, however, significant remaining uncertainty, e.g.,  on DM distribution, which motivates further searches. Direct detection, mainly led by deep underground liquid xenon time projection chambers, has improved sensitivity by two orders of magnitude in the last decade without any DM evidence, see e.g. \cite{Liu:2017drf,Aprile:2017iyp}. While clearly it is too early to abandon the WIMP paradigm, especially in the view of experimental programs in the next five years, the community has started to shift focus to alternative models for DM. 

A particularly interesting, and experimentally largely unexplored region is DM masses at or below the GeV scale. For example, thermal production may also be an attractive option for smaller DM masses \cite{Feng:2008ya}. Other relevant DM models with (sub-)GeV masses include light gravitino DM~\cite{Takayama:2000uz}, inelastic DM \cite{TuckerSmith:2001hy}, light scalar DM \cite{Boehm:2003hm}, or secluded DM \cite{Pospelov:2007mp}. Recently, an anomaly in the absorption profile at 78 MHz in the sky-averaged spectrum \cite{Bowman2018} has been interpreted as an excess cooling of the cosmic gas induced by its interaction with DM particles  lighter than a few GeV \cite{Barkana2018}.

Targets for searches for DM are commonly those of enhanced DM density: the Milky Way galaxy, including  the  GC, dwarf galaxies and groups of galaxies, as well as galaxy clusters. The GC is by orders of magnitude the largest potential source of signal from DM annihilation. Dwarf spheroidal galaxies provide the cleanest target with the potential to derive the DM distribution from spectral velocities and are (unlike the GC) essentially free from conventional sources or diffuse backgrounds that could hamper an identification of DM induced signal. Galaxy clusters are potential targets if a substantial fraction of DM is in substructures. Diffuse backgrounds, such as the Galactic and extragalactic backgrounds, are promising targets, especially exploiting angular autocorrelation or in cross-correlation with other wavelengths, using, for example, galaxy catalogues. For a more detailed review of challenges and opportunities of different gamma-ray signatures and techniques, see e.g. \cite{Conrad:2015bsa,Gaskins:2016cha}.


\section{Scientific requirements} \label{sec:req}

The instrument performance required to achieve the core science objectives, such as the angular and energy resolution, the field of view, the continuum and line sensitivity, the polarization sensitivity, and the timing accuracy, are summarized in Table~\ref{tab:requirements}. 
\begin{itemize}
\item The very large spectral band is required to give a complete view of the main nonthermal processes at work in a given astrophysical object. The 100~keV~--~1 GeV energy band includes, in particular, the 511~keV line from $e^+e^-$ annihilation, the nuclear de-excitation lines, the characteristic spectral bump from pion decay, the typical domains of nonthermal electron bremsstrahlung and Inverse Compton emission, as well as the high-energy range of synchrotron radiation in sources with high magnetic field ($B \ge 1$~G). The  wide energy band is particularly important for the study of blazars, GRBs, Galactic compact binaries, pulsars, as well as the physics of CRs in SNRs and in the ISM. 
\item The wide field of view of the telescope is especially important to enable the measurement of source flux variability over a wide range of timescales both for a-priori chosen sources and in serendipitous observations. Coupled with a sky-scanning mode of operation, this capability enables continuous monitoring of source fluxes that will greatly increase the chances of detecting correlated flux variability with other wavelengths. The  wide field of view is particularly important for the study of blazars, GRBs, Galactic compact objects, supernovae, novae, and extended emissions in the Milky Way (CRs, radioactivity). It will also enable, for example, searches of periodicity and orbital modulation in binary systems. 
\item One of the main scientific requirements is to improve dramatically the detection sensitivity in a region of the electromagnetic spectrum, the so-called MeV domain, which is still largely unknown. The sensitivity requirement is relevant to all science drivers discussed above. Thus, the goal of detecting a significant number ($N > 5$) of SN~Ia in gamma rays after 3 years requires a sensitivity in the 847~keV line $<5 \times 10^{-6}$ ph~cm$^{-2}$~s$^{-1}$  in 1~Ms of integration time (Table~\ref{tab:requirements}). 
\item Another major requirement for a future gamma-ray observatory is to improve significantly the angular resolution over past and current  missions, which have been severely affected by a spatial confusion issue. The required angular resolution will improve $CGRO$/COMPTEL and $Fermi$-LAT by almost a factor of 4 at 1 MeV and 1 GeV, respectively. The targeted angular resolution given in Table~\ref{tab:requirements} is close to the physical limits: for Compton scattering, the limit is given by the Doppler broadening induced by the velocity of the atomic electrons, while for low-energy pair production, the limit is provided by the nuclear recoil. Such an angular resolution will allow a number of currently unidentified gamma-ray sources (e.g. 1323 sources in the 3FGL catalog \cite{4FGL}) to be associated with objects identified at other wavelengths.  
\item The required polarization sensitivity will enable measurements of a gamma-ray polarization fraction $>20$\% in about 40 GRBs per year, and a polarization fraction $>50$\% in about 100 GRBs per year. Such measurements will provide important information on the magnetization and content (leptons, hadrons, Poynting flux) of the relativistic outflows, and, in the case of GRBs at cosmological distance, will address fundamental questions of physics related to vacuum birefringence and Lorentz invariance violation (e.g., \cite{got14}). The  polarization sensitivity will also enable the study of the polarimetric properties of more than 50 blazars, pulsars, magnetars, and black hole systems in the Galaxy.
\item The required spectral resolution for the main science drivers of the mission is largely within the reach of current technologies (Sect. 4). Thus, the main gamma-ray lines produced in SN explosions or by low-energy cosmic ray (LECR) interactions in the ISM are significantly broadened by the Doppler effect, and a FWHM resolution of 3\% at 1 MeV is adequate. In the pair production domain, an energy resolution of 30\% will be more than enough to measure accurately putative spectral breaks and cutoffs in various sources and  distinguish the characteristic pion-decay bump from leptonic emissions. 
\item The required timing performance is mainly driven by the physics of magnetars and rotation-powered pulsars (Sect.~\ref{sec:comp_obj}), as well as by the properties of TGFs (Sect.~\ref{sec:solar}). The targeted microsecond timing accuracy is already achieved in, e.g., the AGILE mission \cite{tav09}. 
\end{itemize}

Requirements for the Ground Segment are standard for an observatory-class mission. Target of Opportunity observations (ToOs) are required to follow particularly important transient events that need a satellite repointing. 
\vskip 2mm
\begin{table}[ht!]
\begin{center}
\caption{\small Estimated number of sources of various classes detectable in 3 years by a gamma-ray mission with the performance shown in Table~\ref{tab:requirements}. The last column gives the expected number of sources not known before in any wavelength. 
\label{tab:nevents}}
\begin{tabular}{| l | l | l |}\hline
\textbf{Source type} & \textbf{Number in 3 yr} & \textbf{New sources}\\ \hline \hline
Galactic & $\sim1000$ & $\sim$400 \\
MeV blazars  & $\sim350$ & $\sim350$ \\
GeV blazars  & {1000 -- 1500} & $\sim350$ \\
Other AGN  ($< $10 MeV)& {70 -- 100} & {35 -- 50}\\
Supernovae  & {10 -- 15} & {10 -- 15}\\
Novae & 4 -- 6 &   4 -- 6 \\
GRBs  & $\sim$700 &  $\sim$700\\ \hline
\textbf{Total} & \textbf{3000 -- 4000} & \textbf{$\sim$1900 (including GRBs)}  \\ \hline
\end{tabular}
\end{center}
\end{table}

Table \ref{tab:nevents} summarizes our estimates of the number of sources detectable in 3 years by a mission having the performance summarized in Table~\ref{tab:requirements}. It is based on current knowledge and $\log N - \log S$ determinations of Galactic and extragalactic sources, including GRBs. It takes information from the latest \textit{Swift}-BAT Hard X-ray survey catalog \cite{Swift}, the \textit{INTEGRAL}-IBIS catalog \cite{ibisc}, and the 4th \textit{Fermi}-LAT catalog \cite{4FGL}. It is noteworthy that the latter catalog contains more than 1300 unidentified sources in the 100 MeV -- 300 GeV range with no counterparts at other wavelength, and most of them will be detected by the new gamma-ray mission, in addition to  a  number of new unidentified sources. The discovery space of such a mission for new sources and source classes would be very large.

\section{Scientific instrumentation}  

\subsection{Technology status}

Astronomy in the MeV range is particularly challenging because mirrors cannot be employed to form images and concentrate the signal. The required performance in terms of field of view and detection sensitivity over a large spectral band (Sect.~\ref{sec:req}) can best be achieved with a relatively heavy instrument (typically 1 ton) that makes optimal use of the physics of gamma-ray interactions with matter. These interactions are dominated by Compton scattering from 100~keV up to about 15 MeV (in silicon), and by electron-positron pair production in the field of a target nucleus at higher energies, with the distinct properties of generating secondary particles and not depositing the photon energy locally, as  happens at lower energies. Both processes can be used for detection by a single instrument made up of two main detectors: a \textbf{Tracker} of secondary particles in which the cosmic gamma rays undergo a Compton scattering or a pair conversion and a \textbf{Calorimeter} to absorb and measure the energy of the scattered gamma rays and electron-positron pairs. In addition, an \textbf{Anticoincidence system} covering the main detectors is needed to veto the prompt-reaction background induced by charged particles in space. 

\begin{figure}[ht]
\centering
\vspace{3mm}
\includegraphics[width=0.95\textwidth]{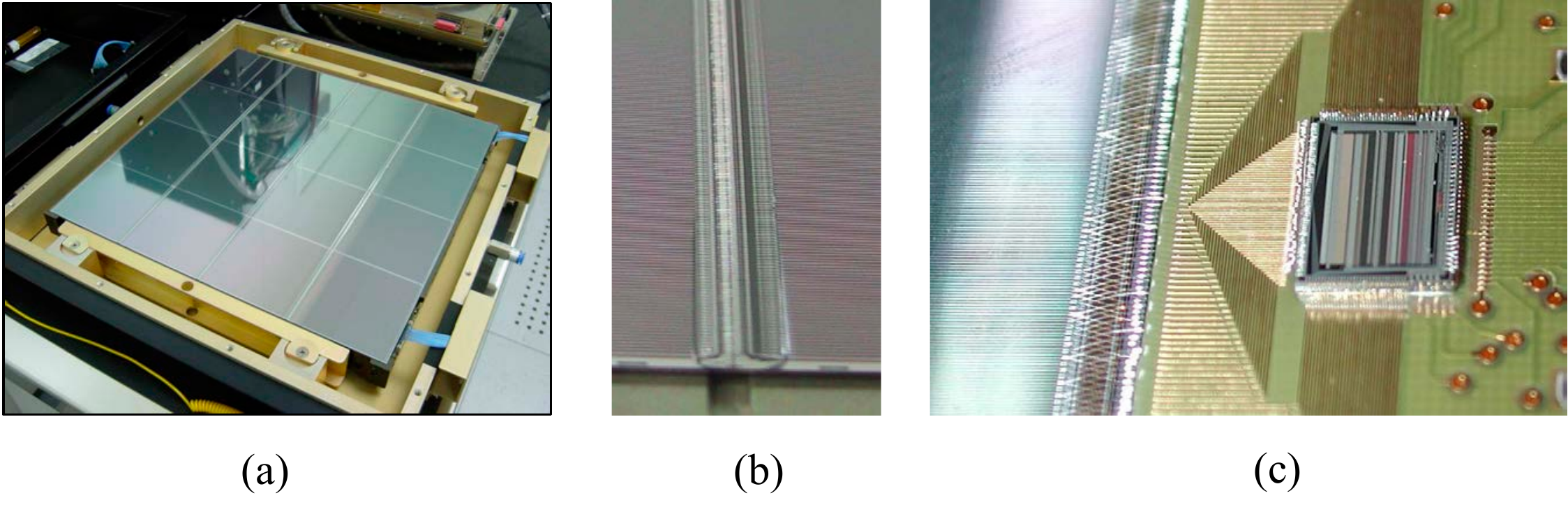}
\caption{{\bf (a)} Assembly of 16 Si microstrip detectors in one layer of a {\it Fermi} LAT tower \cite{atw07}. {\bf (b)} Detail of the {\it Fermi} LAT Tracker showing the wire bonding strip to strip of two Si detectors. {\bf (c)} Detail of the {\it AGILE} Tracker showing the Si sensor bonding with the front-end electronics ASIC through a pitch adaptor (see \cite{tav09} and references therein).}
\label{fig:silicon}       
\end{figure}

Silicon represents the best choice of detector material for the Tracker because of its low atomic number, which favors Compton interactions compared to photoelectric absorption, as well as recent technological advances made on Double-sided Silicon Strip Detectors (DSSDs) and readout microelectronics (see Fig.~\ref{fig:silicon}). In addition, the use of silicon as the scatterer makes it possible to minimize the effect of Doppler broadening, which constitutes an essential physical limit to the angular resolution of a Compton telescope. To increase the detection surface, Si detectors should be daisy-chained with wire bonding strip to strip, each layer comprising typically $4 \times 4$ or $5 \times 5$ DSSDs. 

\begin{figure}[ht]
\centering
\vspace{3mm}
\includegraphics[width=0.85\textwidth]{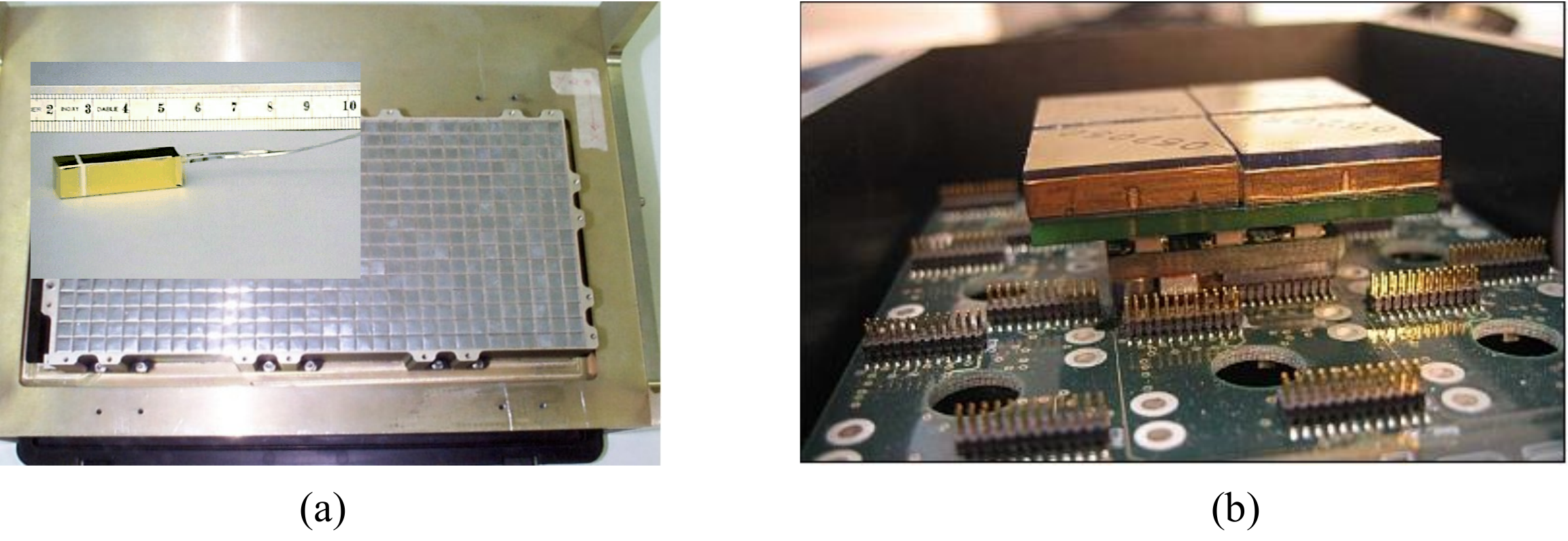}
\caption{{\bf (a)} {\it INTEGRAL}/PICsIT modular detection unit with the inset showing one of the CsI(Tl) bars \cite{lab03}. {\bf (b)} {\it ASIM}/MXGS detector module comprising four $20\times20\times5$~mm$^3$ CZT detectors each having 64 pixels (2.5 mm pixel pitch) \cite{ost19}. MXGS comprises 64 CZT modules providing sensitive area of $1000$~cm$^2$. ASIM was launched on April 2018 and is mounted on the Columbus  module on the International Space Station.}
\label{fig:calorimeter}       
\end{figure}

The basic detector element of the Calorimeter should be made of a material with a high atomic number for an efficient absorption of the scattered gamma rays and the electron-positron pairs. The Calorimeter needs to be a 3-D position-sensitive detector with good energy resolution to capture both Compton and pair interactions, and also contribute efficiently to the background rejection. A pixelated array of high-$Z$ scintillation crystals, such as Thallium activated Cesium Iodine (CsI(Tl)) or Cerium Bromide (CeBr$_3$), readout by silicon drift detectors (SDD) can offer a high stopping power together with good spectral and spatial resolutions. An array of state-of-the-art semiconductors such as CdZnTe (CZT), see Fig.~\ref{fig:calorimeter}, can also provide a very accurate measurement of the interaction location (sub-mm 3D position determination for $E_\gamma>200$~keV), energy determination ($<1\%$ FWHM @ $662$~keV) of the scattered Compton photons and improved polarimetric performance \cite{DTU_3DCZT_1,DTU_3DCZT_2}. 

\begin{figure}[ht]
\centering
\includegraphics[width=0.85\textwidth]{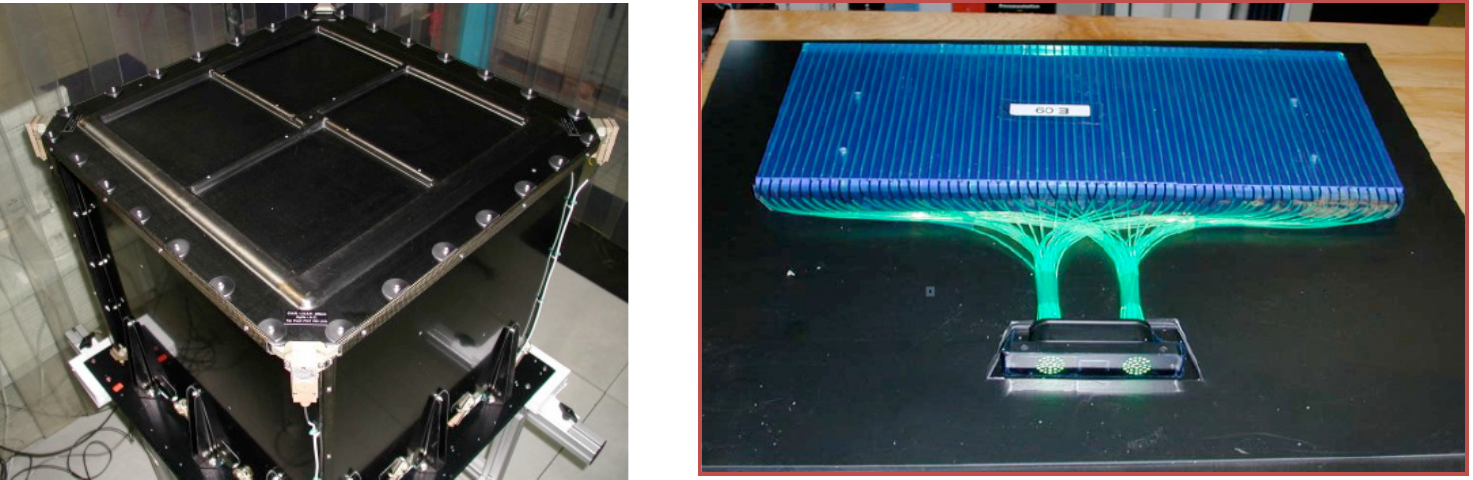}
\caption{{\it Left panel} {\it AGILE} Anticoincidence detector flight unit \cite{per06}. {\it Right panel} Scintillator tile detector assembly (shown unwrapped) of the {\it Fermi}/LAT Anticoincidence \cite{moi07}. The green wavelength-shifting fibers carry light to the optical connector in the foreground.}
\label{fig:acs}       
\end{figure}

The Anticoincidence detector should achieve a charged particle background rejection efficiency $>99.99$\%, which is a standard value already realized in current space experiments such as {\em Fermi}-LAT and {\em AGILE} (Fig.~\ref{fig:acs}). It is classically designed with thin plastic scintillators covering the top and four sides of the instrument. The scintillator tiles can be coupled to silicon photomultipliers (SiPM) by optical fibers, which should provide the best solution to collect the scintillation optical light. 

\subsection{Measurement principle}

\begin{figure}[t]
\centering
\includegraphics[width=0.55\textwidth]{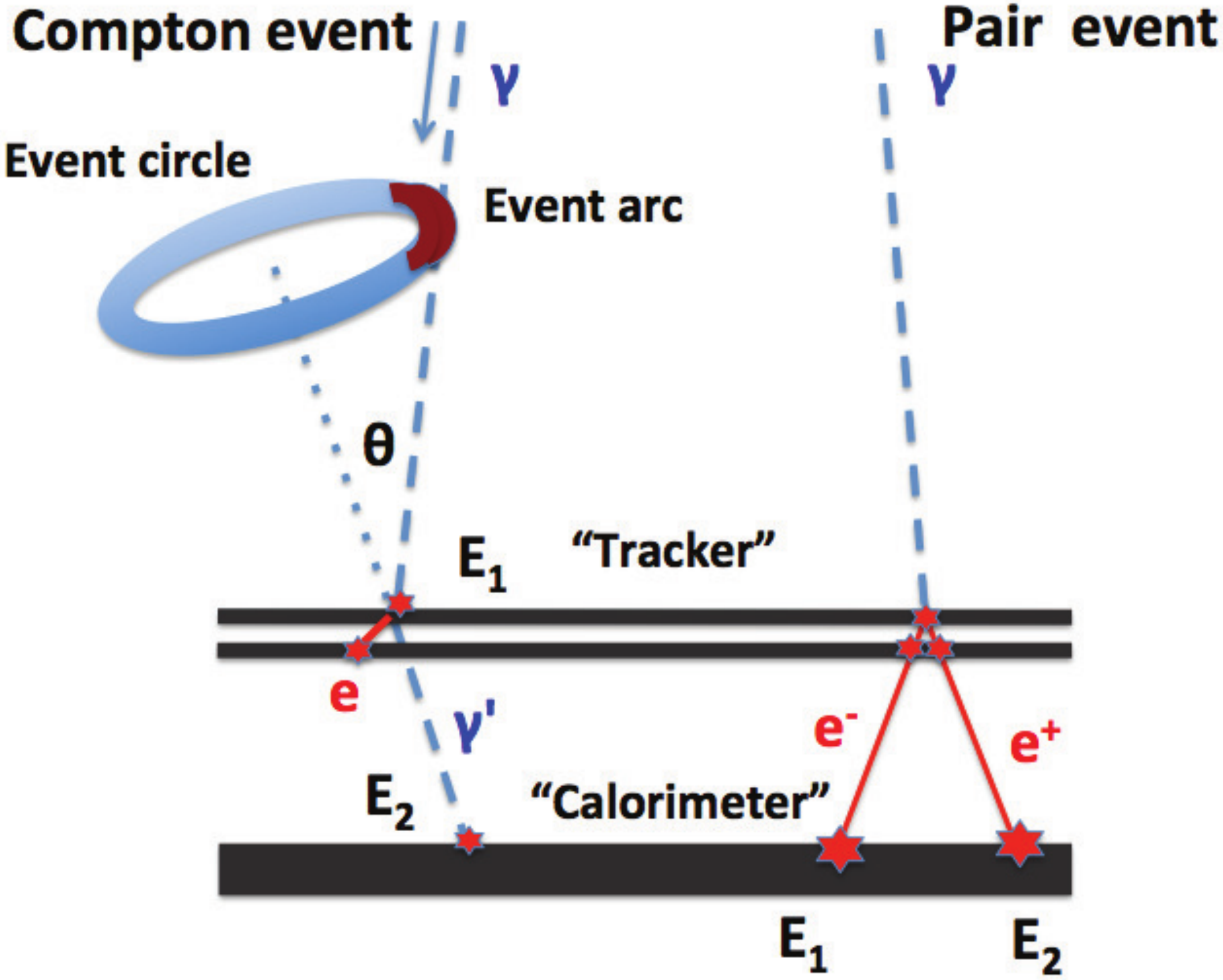}
\caption{Representative  topologies for a Compton event and  for a pair event. Photon tracks are shown in pale blue, dashed, and electron and/or positron tracks are in red, solid.}
\label{fig:evt-top}       
\end{figure}

Figure~\ref{fig:evt-top} shows representative topologies for Compton and pair events. For Compton events, point interactions of the gamma ray in the Tracker and Calorimeter produce spatially-resolved energy deposits, which have to be reconstructed in sequence using the redundant kinematic information from multiple interactions. Once the sequence is established, two sets of information are used for imaging: the total energy and the energy deposit in the first interaction measure the first Compton scatter angle. The combination with the direction of the scattered photon from the vertices of the first and second interactions generates a ring on the sky containing the source direction. Multiple photons from the same source enable a full deconvolution of the image, using probabilistic techniques. For energetic Compton scatters (above $\sim$1 MeV), measurement of the track of the scattered electron becomes possible, resulting in a reduction of the event ring to an arc, hence further improving event reconstruction. Compton scattering angles depend on polarization of the incoming photon, hence careful statistical analysis of the photons for a strong (e.g., transient) source yields a measurement of the degree of polarization of its high-energy emission (e.g. \cite{for08,tat18}).

Pair events produce two main tracks from the created electron and positron. Tracking of the initial opening angle and of the plane spanned by the electron and positron tracks enables direct back-projection of the source position. Multiple scattering of the pair in the tracker material (or any intervening passive materials) leads to broadening of the tracks and limits the angular resolution. The nuclear recoil taking up an unmeasured momentum results in an additional small uncertainty. The energy of the gamma ray is measured using the Calorimeter and information on the electron and positron multiple scattering in the Tracker. Polarization information in the pair domain is given by the azimuthal orientation of the electron-positron plane.

\subsection{Artificial intelligence for event reconstruction}
The high rate of charged particle and gamma-ray background and the limited bandwidth in data transmission to ground requires a multi-level trigger and data selection scheme. At trigger level, charged particles can be removed by a plastic scintillator veto with high efficiency. Taking into account time-of-flight information helps to keep efficiency high for the detection of high-energy pair events that leak out of the calorimeter.
The expected pretrigger rate for an e-ASTROGAM-like detector on a low inclination LEO is approximately $65$\,kHz. After trigger level, this rate reduces to a total rate of around $4$\,kHz \cite{eastrogam}.
In the next data reduction  step, smart and fast event selection is required. 
Data processing on board a satellite is constrained by computational resources and communication bandwidth. The former limits complexity of a processing chain, the latter sets an upper limit on the amount of data that can be transmitted to Earth and thereby sets a lower bound for the amount of selection required to extract the relevant data. 
One straightforward way to raise the efficiency of on-board processing is to ensure the correct categorization of an event at the beginning of the processing chain. This is a task where \emph{Machine Learning} can play a major role. Since the tracker data has the largest discrimination power concerning the event type (Compton- or Pair-event) any classification attempt should start there. The x- and y-strips in each tracker layer provide a natural way to generate images from the event using raw data (i.e., x-z- and y-z-maps in ADC channels). This makes the application of image recognition techniques feasible.  Convolutional Neural Nets (CNN) \cite{LecunCnn} are the leading technique for this task.
However the computational effort for processing an image via CNNs rises with the size of the image. The capability to run a complex CNN in an FPGA-accelerated 
System-on-Chip  environments was demonstrated by 
e.g.\,\cite{Zynqnet}.  There are also efforts to develop ASIC solutions for CNNs (e.g.\,\cite{SCNN}) that provide an increase in speed.
Such a network could tag each event for further processing or disposal. 

In a further step, pair events are processed by an on-board Kalman filter
(e.g.\,\cite{AGILE:Kalman}) to check the event for viability. A natural representation of a Kalman filter in the context of Deep Learning is a Recurrent Neural Net (RNN) \cite{MirowskiLecun}. In a small toy example, Gu et al.\,\cite{Gu} show that the use of an RNN leads to a model that behaves equivalently to a Kalman filter but possesses a better resistance to noisy input than the conventional approach. 
As above, further developments are required to make such systems operate on space-grade processors.

\section{Mission profile} \label{sec:mission}

Previous studies, in particular for  e-ASTROGAM (a proposal for ESA's M5 opportunity), have shown that the scientific requirements presented in Sects.~\ref{sec:science} and \ref{sec:req} could be met by a M-size mission. The typical envelopes of the mission are: 
\begin{itemize}
    \item Payload mass: about 1 ton
    \item Satellite dry mass: about 2.5 tons
    \item Satellite power: about 2 kW in nominal science operation
    \item Telemetry budget: about 1.5 Mbps
\end{itemize}

The detection sensitivity requirement (Sect.~\ref{sec:req}) would be consistent with the launch of the mission to an equatorial low-Earth orbit (LEO) (typical inclination $i<2.5^\circ$ and eccentricity $e<0.01$) of altitude in the range 550~--~600 km.  Such an orbit is preferred for a variety of reasons. It has been demonstrated to be only marginally affected by the South Atlantic Anomaly and is therefore a low-particle background orbit, ideal for high-energy observations. The orbit is practically unaffected by precipitating particles originating from solar flares, a virtue for background rejection. Finally, ESA has satellite communication bases near the equator that can be efficiently used as mission ground stations.

Extensive simulations of the detection performance using state-of-the-art numerical tools \cite{zog06,bul12} and a detailed numerical mass model of the satellite together with a thorough model for the background environment have shown that a mission like e-ASTROGAM would achieve \cite{tat16}: 
\begin{itemize}
\item Broad energy coverage ($\sim$0.1 MeV to 1 GeV), with nearly two orders of magnitude improvement of the continuum sensitivity in the range 0.1 -- 100 MeV compared to previous missions;
\item Excellent sensitivity for the detection of key gamma-ray lines e.g. sensitivity for the 847~keV line from thermonuclear supernovae 70 times better than that of the {\it INTEGRAL} spectrometer (SPI);
\item Unprecedented angular resolution both in the MeV domain and above a few hundreds of MeV,  i.e., improving the angular resolution of the COMPTEL telescope on board the CGRO and that of the {\it Fermi}/LAT instrument by a factor of $\sim$4 at 5 MeV and 1~GeV, respectively (e.g. the e-ASTROGAM Point Spread Function (68\% containment radius) at 1 GeV is 9').
\item Large field of view ($>$ 2.5 sr), ideal to detect transient Galactic and extragalactic sources, such as X-ray binaries and gamma-ray bursts;
\item Pioneering polarimetric capability for both steady and transient sources \cite{tat18}. 
\end{itemize}

\section{Technology readiness and foreseen developments}

Considering the ASTROGAM concept, the detector technology, silicon tracker, plastic scintillator-based anticoincidence, and crystal calorimeter have been already successfully used in space, and the payload would be based on a very high (TRL) for all crucial detectors and associated electronics. However, a moderate R\&D effort should be considered.

For the silicon tracker, the 2D bonding of 4x4 (or even 5x5) DSSDs needs some R\&D activities to implement mechanical jigs to guarantee the alignment of silicon tiles during the bonding on the two sides of silicon planes. In addition, a bonding machine, able to work on a large area of such silicon planes, should be identified on the market. The 2D bonding procedure has already been established for the PAMELA \cite{pamela} and AMS \cite{ams} space missions and it is well established. The current fabrication technology of large silicon wafers up to 300 mm in diameter could be also investigated to reduce the number of bondings by using larger area of DSSDs.

For the calorimeter, the technology and fabrication process of the Silicon Drift Detectors is the same as the one that was the subject of an extensive development activity within the assessment phase of the LOFT ESA M3 mission \cite{loftass} and more recently for the eXTP project \cite{extp}. The low-noise front-end electronics would require some efforts to optimize the signal to noise performance and to reduce the power consumption.

A beam test campaign would also need to study the performance of the single detector and the whole system as well. In addition, environmental space testing would be also required to space qualify the assembling of the detectors.

Finally, it is not unlikely that by the mid-XXI century new technologies are available and ready for space missions, from monolithic Si tracking elements to completely new concepts.


\section{Conclusions}

The e-ASTROGAM  concept for a gamma-ray space observatory  can revolutionize  the  astronomy  of  medium/high-energy  gamma  rays  by  increasing  the  number of  known  sources  in  this  field  by  more  than  an  order  of  magnitude  and  providing  polarization  information  for  many  of  these  sources.

The technology is ready but foreseen improvements in the next decade, in a framework called ASTROMEV, can further enhance the performance. ASTROMEV will be a protagonist of multi-messenger astronomy and play a major role in the development
of time-domain astronomy. New  windows of opportunity (sources of gravitational waves, neutrinos, ultra high-energy cosmic rays) will be fully and uniquely explored.

\end{document}